\documentclass[a4paper,11pt]{article}
\usepackage{pos}
\makeatletter
\g@addto@macro\bfseries{\boldmath}
\makeatother

\title{{\bf Summary of Working Group 4:
Mixing and mixing-related CP
violation in the $B$ system: $\mathbf{\Delta m}$, $\mathbf{\Delta \Gamma}$, $\mathbf{\phi_{s}}$, $\mathbf{\phi_{1}/\beta}$, $\mathbf{\phi_{2}/\alpha}$,  $\mathbf{\phi_{3}/\gamma}$}}
\ShortTitle{CKM 2021: Summary of Working Group 4}

\author[a]{Veronika Chobanova}
\author[b]{Matthew Wingate}
\author[c]{Yosuke Yusa}
\author[a]{Jeremy Dalseno}
\author[d,e]{Kristof De Bruyn}
\author[f]{Ulrik Egede}
\author[g]{Fabio Ferrari}
\author[h]{Thibaud Humair}
\author[i]{Anna Lupato}
\author[e]{Eleftheria Malami}
\author[j]{Wenbin Qian}
\author[a]{Ram\'{o}n \'{A}ngel Ruiz Fern\'{a}ndez}
\author[k]{Vladyslav Shtabovenko}
\author[l]{Justus Tobias Tsang}
\author[m]{Luiz~Vale~Silva}

\affiliation[a]{Instituto Galego de F\'{i}sica de \'{A}ltas Enerx\'{i}as, Universidade de Santiago de Compostela,
  Santiago de Compostela, Spain}

\affiliation[b]{DAMTP, University of Cambridge,
Cambridge, United Kingdom}

\affiliation[c]{Department of Physics, Nigata University,
Nigata, Japan}

\affiliation[d]{Van Swinderen Institute for Particle Physics and Gravity, University of Groningen,
9747 Groningen, Netherlands}

\affiliation[e]{Nikhef,
Science Park 105, NL-1098 XG Amsterdam, Netherlands}

\affiliation[f]{School of Physics and Astronomy, Monash University, Melbourne, Australia}

\affiliation[g]{University of Bologna and INFN, via Irnerio 46, Bologna, Italy}

\affiliation[h]{Max Planck Institute for Physics,  Föhringer Ring 6, 80805 München, Germany}

\affiliation[i]{The University of Manchester,
 Manchester , United Kingdom}
 
 \affiliation[j]{University of Chinese Academy of Sciences}

\affiliation[k]{Institut für Theoretische Teilchenphysik (TTP), Karlsruhe Institute of Technology (KIT), Wolfgang-Gaede-Straße 1, 76131 Karlsruhe, Germany}

\affiliation[l]{IMADA \& CP$^3$-Origins. University of Southern Denmark.
  Campusvej 55, DK-5230 Odense, Denmark}

\affiliation[m]{Instituto de F\'{i}sica Corpuscular, Universitat de Val\`encia - Consejo Superior de Investigaciones Cient\'{i}ficas,
  Parc Cient\'{i}fic, E-46980 Paterna, Valencia, Spain}

\emailAdd{veronika.chobanova@cern.ch}
\emailAdd{M.Wingate@damtp.cam.ac.uk}
\emailAdd{yusa@hep.sc.niigata-u.ac.jp}
\emailAdd{jeremy.peter.dalseno@cern.ch}
\emailAdd{kristof.debruyn@cern.ch}
\emailAdd{ulrik.egede@monash.edu}
\emailAdd{fabio.ferrari@cern.ch}
\emailAdd{anna.lupato@cern.ch}
\emailAdd{emalami@nikhef.nl}
\emailAdd{wenbin.qian@ucas.ac.cn}
\emailAdd{ramon.angel.ruiz.fernandez@cern.ch}
\emailAdd{v.shtabovenko@kit.edu}
\emailAdd{tsang@imada.sdu.dk}
\emailAdd{luizva@ific.uv.es}
\abstract{
This is a summary of the latest results in $B$ meson mixing and mixing-related {\it CP} violation presented at
CKM 2021. We place these in the context of both recent experimental measurements, theoretical developments and future prospects in the field.}

\FullConference{%
  11th International Workshop on the CKM Unitarity Triangle (CKM 2021)\\
  22-26 November 2021\\
  University of Melbourne, Australia (virtual)
}

\begin{document}
\maketitle

\section{Introduction}

The study of neutral $B$ meson oscillations provides many insights into quark flavour dynamics.
The oscillation frequencies can be calculated precisely within the Standard Model (SM), and any experimentally measured deviation could hint at contributions from New Physics (NP).  Standard Model predictions indicate that \textit{CP} violation in mixing should be small, so discovery of large violation would also herald New Physics.

This summary reviews contributions to the CKM 2021 workshop in Working Group 4. In particular, theoretical and experimental progress determining $B$ meson mixing properties are discussed. New measurements of the CKM angle $\gamma$, the $B_s$ mass difference, and time-dependent \textit{CP} violation parameters are presented. Some puzzles in the $B_s^0 \to D_s^\mp K^\pm$ system are summarized, and new ideas for $\phi_2(\alpha)$ measurements are presented.  Consequences for global CKM fits are reviewed.  Finally, an outlook is given as to what future upgrades and experiments can bring us.

\section{Standard Model predictions of $B$ meson mixing parameters}
\label{sec:Bmixing_theory}

The $B_{d,s}^0 - \overline{B}_{d,s}^0$ systems each feature three physical observables that are of great interest for probing our understanding of the Standard Model as well as looking for New Physics.
These are the mass differences $\Delta m_{d,s}$, the width differences $\Delta \Gamma_{d,s}$, and {\it CP} asymmetries in flavor-specific decays $a^{d,s}_{\textrm{fs}}$. A statistically significant discrepancy between theory predictions and experimental measurements for $\Delta m_{d,s}$ would be a hint for NP effects entering via heavy particles in the loops. Light beyond Standard Model (BSM) particles feebly interacting with SM fields can instead induce a disagreement between theory and experiment in the case of $\Delta \Gamma_s$. The most recent experimental values for these observables read
\begin{alignat}{2}
    \Delta m^{\rm exp}_d &=  (0.5065 \pm 0.0019)
	\; \text{ps$^{-1}$} &  \quad \text{\cite{Amhis:2019ckw}}\,, \\
	\Delta m^{\rm exp}_s &=  (17.765 \pm 0.006)
	\; \text{ps$^{-1}$} & \quad \text{\cite{Amhis:2019ckw}}\,, \\
	\Delta \Gamma^{\rm exp}_s  &=  (0.082 \pm 0.005)\;
	\text{ps}^{-1} %\qquad \hspace*{1.5em}
	& \quad  \text{\cite{Amhis:2019ckw}} \label{eq:exp}\,.
\end{alignat}
The width difference $\Delta \Gamma_d$ is still not measured \cite{Amhis:2019ckw}.

Since the $W$ boson is so massive compared to hadronic scales, neutral meson mixing of $B_q$ ($q=d,s$) mesons can be parameterised in terms of matching coefficients in effective Hamiltonians
and non-perturbative matrix elements. The non-perturbative matrix elements of the five parity-even, dimension-6, $\Delta B = 2$ operators can be computed from first principles by means of lattice QCD
simulations. These are often cast in the form of bag parameters, i.e.\ are
normalised by their vacuum saturation values. The first of these operators, $O_1 \propto (\bar{b}\gamma_\mu q)_L(\bar{b}\gamma^\mu q)_L$, is of
particular relevance since the experimentally measurable mass difference
$\Delta m_q$ ($q=l,s$) can be parameterised as
\begin{equation}
  \Delta m_q = \left|{V_{td} V_{tq}^*}\right|^2 \, \mathcal{K} \, M_{B_q} \,f^2_{B_q} \hat{B}^{(1)}_{B_q}\,,
\end{equation}
where the $\mathcal{K}$ factor is known and independent of the light quark. 
Precise knowledge of the
nonperturbative decay constant $f_{B_q}$ and the bag parameter $\hat{B}^{(1)}_{B_q}$,
in combination with the experimental measurements, therefore allows the
extraction of the CKM matrix elements $|V_{tq}|$. The SU(3) breaking ratio $\xi^2
= (f^2_{B_s} \hat{B}^{(1)}_{B_s})/(f^2_{B_d} \hat{B}^{(1)}_{B_d})$ gives access
to the ratio $\left|V_{td}/V_{ts}\right|$~\cite{Bernard:1998dg}.

Since the last meeting of this workshop, CKM 2018, where the results from the
Fermilab/MILC collaboration~\cite{FermilabLattice:2016ipl} were discussed, new
computations by HPQCD~\cite{Dowdall:2019bea} and by
RBC/UKQCD~\cite{Boyle:2018knm} have become available. These were reviewed at CKM 2021 by Tsang~\cite{Tsang:2022}. Both of these works
include ensembles with physical pion mass ensembles, removing the need for a
chiral extrapolation and thereby eliminating an important source of systematic
uncertainty. The large bottom quark mass can cause sizeable discretisation effects
in lattice QCD simulations. To circumvent this, the heavy quark can be simulated
with an effective action which enables direct simulations at the bottom quark
mass in exchange for difficult-to-reduce systematic
uncertainties. Alternatively, simulations can take place at lighter than
physical heavy quark masses without the need for an effective action, but
instead requiring an extrapolation to the physical $b$-quark mass. 

The HPQCD result~\cite{Dowdall:2019bea} follows the first approach with the heavy quarks discretised using the non-relativistic QCD
action~\cite{Lepage:1992tx}. The light quarks use the highly improved staggered quark (HISQ) 
action, a successor to the AsqTad staggered action used in Ref.~\cite{FermilabLattice:2016ipl}.
The matrix elements of all five operators are presented for $B_d$ and $B_s$ mixing. The
individual bag parameters have uncertainties ranging between 4 and 8\% improving
on, and in agreement with, the results from
Fermilab/MILC~\cite{FermilabLattice:2016ipl}. Due to partial cancellations of
statistic and systematic uncertainties, the uncertainty for the ratio of bag
parameters is reduced to approximately 2.5\%.

The RBC/UKQCD~\cite{Boyle:2018knm} collaboration uses the chirally symmetric
domain wall fermion action for all quark flavours. Data are simulated for quark
masses from below the charm quark mass to approximately half the bottom quark
mass. Suitable SU(3) breaking ratios such as $\xi$ and the ratio of bag
parameters are formed because their benign heavy quark dependence allows for a
controlled extrapolation to the physical bottom quark mass. The ratios of bag
parameters and the phenomenologically interesting quantity $\xi$ are determined
around the percent level. The dominating systematic uncertainty stems from the
extrapolation in the heavy quark mass. This source of uncertainty can be
systematically improved by including ensembles with finer lattice spacings which
allow to simulate closer to the physical $b$-quark mass. Efforts to achieve this
and to extend the analysis to the full five-operator basis are in progress. First
results for a joint analysis between RBC/UKQCD and JLQCD have been reported
recently~\cite{Boyle:2021kqn}.

The results~\cite{FermilabLattice:2016ipl,Dowdall:2019bea,Boyle:2018knm} are
highly complementary as they utilise very different methodologies. In particular
the gauge field ensembles, the light quark action, and the heavy quark actions
are all different between the three works. The obtained results are mutually
compatible, but the extraction of $\left|V_{td}\right|$, $\left|V_{ts}\right|$ 
and their ratio remains dominated by theoretical uncertainties, necessitating 
further, more precise predictions.

In addition to the lattice results for the dimension-6 operator matrix elements, 
two recent, related works are noteworthy. The first is a set of sum rule
calculations of the dimension-6 operator matrix elements~\cite{King:2019lal}. These are in good
agreement with the aforementioned lattice results \cite{FermilabLattice:2016ipl,Dowdall:2019bea}.
The second is a
lattice computation of the dimension-7 operators contributing at next-to-leading order in $1/m_b$ to the $B_s - \overline{B}_s$ width difference $\Delta \Gamma_s$~\cite{Davies:2019gnp}. This is the first time that 
$\Delta \Gamma_{1/m_b}$ has been determined using lattice calculations, replacing rough estimates from the vacuum saturation approximation.  The theory uncertainty coming from $\Delta \Gamma_{1/m_b}$ is still one impediment to matching the experimental precision. 

The theoretical prediction for $\Delta \Gamma_s$ is also known to be affected by large perturbative uncertainties. These uncertainties quantify uncalculated two- and three-loop QCD corrections to the Wilson coefficients obtained from the matching between $|\Delta B|=1$ and $|\Delta B|=2$ effective Hamiltonians. 
Different contributions to these matching coefficients can be enumerated in terms of different operator insertions on the $|\Delta B|=1$ side. These comprise two insertions of an operator from the following categories: current-current $Q_{1-2}$, four-fermion penguin $Q_{3-6}$ and chromomagnetic penguin
$Q_{8}$. 

The current state-of-the-art prediction for $\Delta \Gamma_s$ corresponds to NLO accuracy 
\cite{Beneke:1998sy,Beneke:1996gn,Ciuchini:2003ww,Beneke:2003az,Lenz:2006hd} with partial NNLO results \cite{Asatrian:2017qaz,Asatrian:2020zxa}. The latter, however, include only fermionic contributions proportional to the number of flavors $N_f$. At this workshop, Shtabovenko reported on the main steps towards the full NNLO prediction. All contributions $Q_i \times Q_j$ had been calculated at two-loop accuracy and the three-loop current-current $Q_{1-2} \times Q_{1-2}$ piece had been evaluated. For simplicity, the Wilson coefficients were calculated as an expansion in $z\equiv m_c^2/m_b^2$ up to $\mathcal{O}(z)$.
Furthermore, the two-loop double chromomagnetic penguin insertion $Q_{8} \times Q_{8}$ is formally of $\textrm{N}^3$LO. The analytic two-loop results can be found in \cite{Gerlach:2021xtb,Gerlach:2022wgb},
while the three-loop calculation has just appeared \cite{Gerlach:2022hoj}.

By including all the two-loop corrections at one's disposal, we now have new theory predictions for the ratio $\Delta \Gamma_s / \Delta m_s$, which has the nice feature of being
independent of $V_{ts}$ so that the results are not affected by the existing $V_{cb}$ controversy.
Shtabovenko and collaborators find
\begin{align}
	\frac{\Delta\Gamma_s}{\Delta m_s} 
	&=
	(4.70 
	{}^{+0.32}_{-0.70}{}_{\rm scale}
	\pm 0.12_{B\tilde{B}_S}
	\pm 0.80_{1/m_b}
	\pm 0.05_{\textrm{input}})
	\times 10^{-3}
	\quad
	(\textrm{pole})\,,  \nonumber\\
	\frac{\Delta\Gamma_s}{\Delta m_s} 
	&=  
	(5.20
	{}^{+0.01}_{-0.16}{}_{\rm scale}
	\pm 0.12_{B\tilde{B}_S}
	\pm 0.67_{1/m_b}
	\pm 0.06_{\textrm{input}}) \times 10^{-3} \quad
	(\overline{\textrm{MS}})\,, 
	\label{eq::dGdM}
\end{align}
where ``scale'' describes the uncertainties related to the choice of the renormalization scale, 
while ``$B\tilde{B}_S$'' is related to the variations of the leading order bag parameters
and ``input'' is linked to the uncertainties in the values of the strong coupling constant, CKM parameters and the quark masses.
Finally, ``$1/m_b$'' denotes the uncertainties from the power-suppressed corrections in
the operator product expansion. The schemes ``$\overline{\text{MS}}$'' and ``pole'' refer to the way how one treats the $m_b^2$ prefactor in the theoretical formula for $\Delta \Gamma_s$ (c.f. \cite{Gerlach:2021xtb,Gerlach:2022wgb} for more details). One can choose it as an $\overline{\text{MS}}$ or an on-shell mass. Notice that in both schemes all masses apart from this $m_b^2$ prefactor are always treated in the $\overline{\text{MS}}$ scheme.
A numerical update of these numbers featuring the three-loop current-current contributions is currently in preparation and will finally allow a prediction of $\Delta \Gamma_s$ at NNLO accuracy.

\section{Measurements of $B$ meson lifetime and mixing properties, and time-dependent CPV at the LHC}
The LHC experiments ATLAS, CMS and LHCb reported recent measurements of the three CKM angles, the {\it CP}-violating phase $\phi_{s}$, the $B_{s}^{0}$ lifetime properties and the mixing parameter $\Delta m_{s}$. Most notably, the combined LHCb precision in $\Delta m_{s}$~\cite{LHCb:2020qag} is three times better than the previous world average~\cite{Amhis:2019ckw}. Furthermore, LHCb made the first time-dependent observation of {\it CP} violation in the $B_{s}^{0}$ system, in the $B_{s}^{0}\to KK$ decay mode~\cite{alpha}. The updates on $\phi_{s}$ include both a significant increase in the experimental precision as well as a deeper understanding in the penguin contributions from global fits to signal and control modes.

\subsection{Time-dependent measurement of $\gamma$ and $\Delta m_s$ at LHCb}

\begin{figure}
    \centering
    \includegraphics[width=0.6\linewidth]{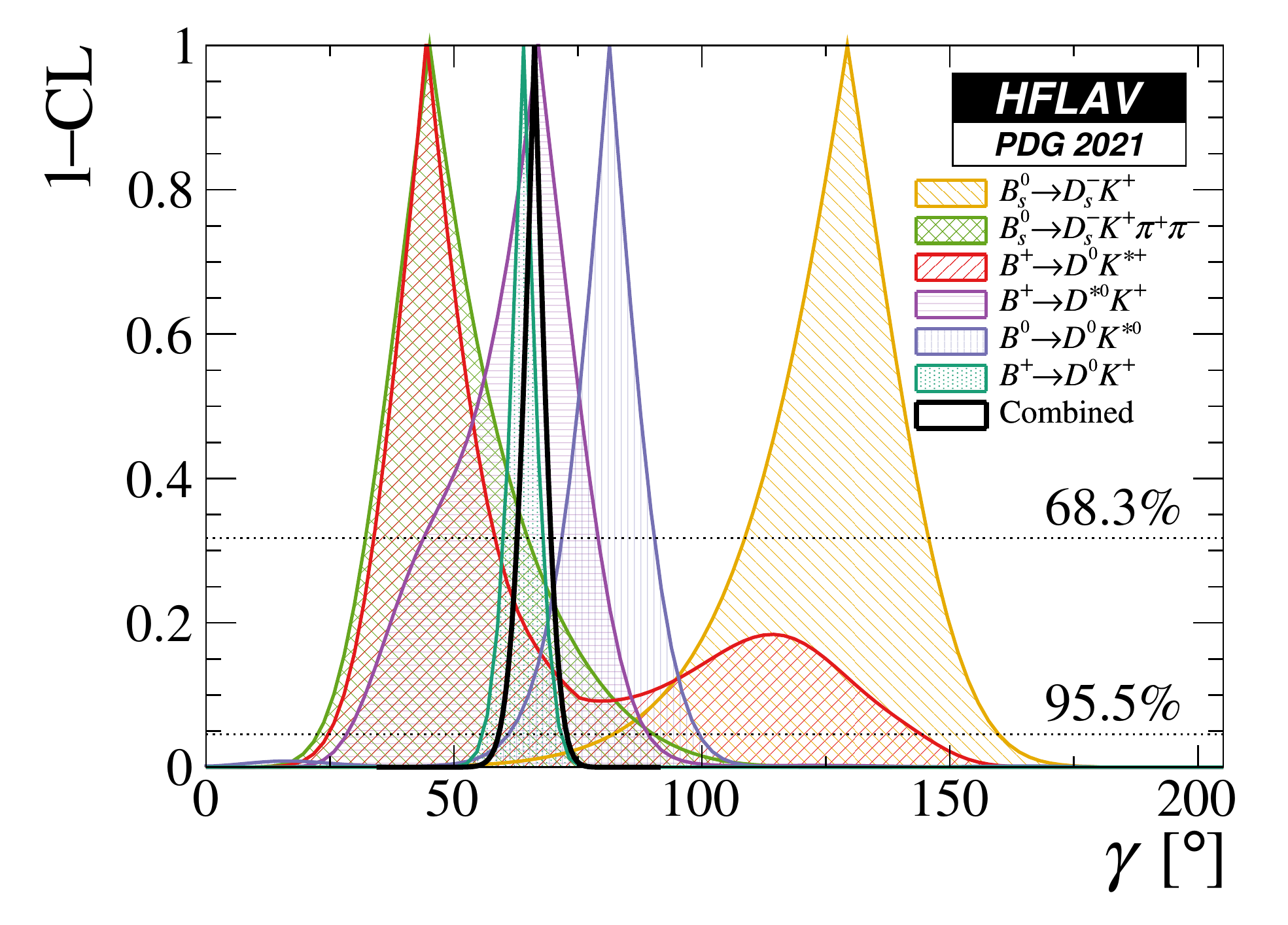}
    \caption{The confidence level for the measurements of the CKM angle $\gamma$ broken down by decay mode~\cite{Amhis:2019ckw}. The new measurement~\cite{LHCb:2020qag} using $B^0_s \to D_s^{\mp}K^{\pm}\pi^{\pm}\pi^{\mp}$ is hatched green.}
    \label{fig:gamma}
\end{figure}

The most recent CKM angle $\gamma$ measurements performed at LHCb were presented, including the first measurement that takes advantage of mixing-induced {\it CP} violation between $B^0_s \to D_s^{\mp}K^{\pm}\pi^{\pm}\pi^{\mp}$ and $\overline{B}^0_s \to D_s^{\pm}K^{\mp}\pi^{\mp}\pi^{\pm}$ decays. The decays are reconstructed in proton-proton collision data corresponding to an integrated luminosity of 9 fb$^{-1}$ recorded with the LHCb detector at a centre-of-mass energy of 13~TeV. 
In these decays, the sensitivity to the weak phase results from the interference between b $\to$ c and b $\to$ u transitions achieved through $B^0_s - \overline{B}^0_s$ mixing. Mesons comprising a beauty quark and a strange quark can oscillate between particle and antiparticle flavour eigenstates, with a frequency given by the mass difference between heavy and light mass eigenstates, $\Delta m_s$. Due to the interference between mixing and decay amplitudes, the physical {\it CP}-violating observables in these decays are functions of a combination of $\gamma$ and the mixing phase $\beta_s$. To account for the non-constant strong phase across the phase space, one can either perform a time-dependent amplitude fit or select a suitable phase-space region and introduce a coherence factor as additional hadronic parameter to the decay-time fit. Both approaches are explored. A time-dependent amplitude analysis is performed to extract the {\it CP}-violating weak phase $\gamma - 2\beta_s$, and subsequently $\gamma = \left(44 \pm 12\right)^\circ$ modulo 180$^\circ$, where statistical and systematic uncertainties are combined \cite{LHCb:2020qag}. An alternative model-independent measurement, integrating over the five-dimensional phase space of the decay, yields $\gamma = \left(44^{+20}_{-13}\right)^\circ$ modulo 180$^\circ$ \cite{LHCb:2020qag}. Therefore, a good agreement between the two methods has been achieved. As shown in Fig.~\ref{fig:gamma}, this new result has been combined with the previous results of $\gamma$ within the framework of HFLAV~\cite{Amhis:2019ckw}. While the combination is dominated by the resolution obtained for the various $B^+ \to D^0 K^+$ modes, the new result has comparable precision to other modes and adds to the overall consistency in the determination of the angle. The world average using all measurements is $\gamma = \left(66.2^{+3.4}_{-3.6}\right)^\circ$.

LHCb also reported two new measurements of $\Delta m_s$. 
The $B^0_s$ - $\overline{B}^{0}_s$ oscillation frequency is measured from the flavour-specific channel 
$B^0_s \to D_s^{-}\pi^{+}\pi^{+}\pi^{-}$ to be $\Delta m_s = (17.757 \pm 0.007($stat$)\pm0.008($syst$))$ ps$^{-1}$ \cite{LHCb:2020qag}. Moreover, LHCb presented the measurement of $\Delta m_s$ through the $B^0_s \to D_s^{-}\pi^{+}$ decay channel \cite{LHCb:2021moh}. This measurement improves upon the current $\Delta m_s$ precision by a factor of two and is found to be $\Delta m_s = (17.7683 \pm 0.0051($stat$)\pm0.0032($syst$))$ ps$^{-1}$. Combining all LHCb $\Delta m_s$ measurements, the average is $\Delta m_s = \left(17.7656 \pm 0.0057\right)$~ps$^{-1}$~\cite{LHCb:2021moh}. This value is compatible with, and considerably more precise than, the predicted value from lattice QCD \cite{Dowdall:2019bea} and sum rule \cite{King:2019lal} calculations  of $ 18.4^{+0.7}_{-1.2}$ ps$^{-1}$ \cite{DiLuzio:2019jyq}.

\subsection{Measurement of CPV in $B\to hh^{(')}$ and $B^0\to D^*D$ decays at LHCb}
\begin{figure}[h!]
    \centering
    \includegraphics[width=0.6\linewidth]{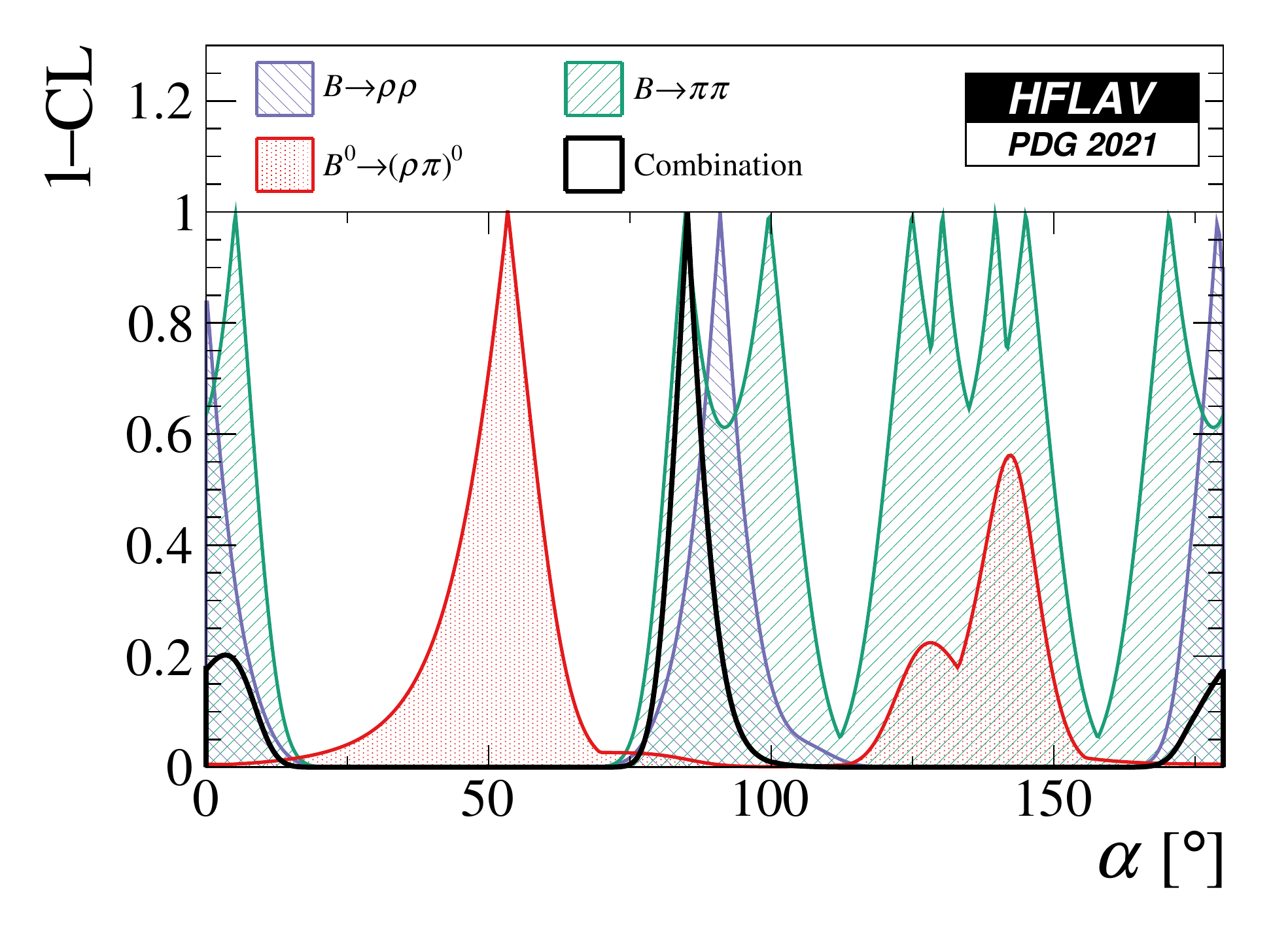}
    \caption{World average of $\alpha$ (black line) and for the individual decay modes (coloured lines)~\cite{Amhis:2019ckw}.}
    \label{alpha-figure}
\end{figure}

The LHCb collaboration reported measurements of time-dependent {\it CP} asymmetries of $B^0 \rightarrow \pi^+\pi^-$ and $B^0_s\rightarrow K^+K^-$ and of time-integrated {\it CP} asymmetries in $B^0\rightarrow K^+\pi^-$ and $B_s^0\rightarrow K^-\pi^+$ decays~\cite{alpha}. These are based on the $pp$ collisions collected during Run~2 of the LHC in 2015 and 2016 and correspond to an integrated luminosity of 1.9 $\mathrm{fb^{-1}}$ at a centre of mass energy of $13$ TeV. The measurements are combined with Run~1 results~\cite{hh-Run1}, yielding, in $B^0_s\rightarrow K^+K^-$,
 $C_{KK} = 0.172\pm 0.031$, $S_{KK} = -0.139\pm 0.032$ and $A^{\Delta \Gamma}_{KK} = -0.897 \pm 0.087$.
The combined time-dependent {\it CP} asymmetry in $B^0_s\rightarrow K^+K^-$ decays of the parameters $C_{KK}$, $S_{KK}$, $A^{\Delta \Gamma}_{KK}$ is the first observation of time-dependent {\it CP} violation in the $B_s^0$ system, excluding the hypothesis of {\it CP} conservation by more than $6.5\sigma$;
in $B^{0}\to \pi^{+}\pi^{-}$, $C_{\pi\pi} = -0.320 \pm 0.038$ and $S_{\pi\pi} = -0.672\pm0.034$; in $B^{0}\to K^{+}\pi^{-}$ and $B^{0}_{s}\to K^{-}\pi^{+}$, $A_{CP}(B^{0})=0.0831\pm0.0034$ and $A_{CP}(B^{0}_{s})=-0.225\pm0.012$. 
The measurements of $C_{\pi\pi}$, $S_{\pi\pi}$, $A_{CP}(B^{0})$ and $A_{CP}(B_s^0)$ are the most accurate to date and are compatible with previous results provided by the B-factories~\cite{babarhh, bellehh}. 
The world average of $\alpha$ as calculated by HFLAV in an isospin analysis of time-dependent {\it CP}-violating parameters in $B^{0} \to \pi^+\pi^-,  \rho\pi, \rho\rho$ decays is shown in Fig.~\ref{alpha-figure}. It yields  $\left(85.4^{+4.8}_{-4.3}\right)^{\circ}$~\cite{Amhis:2019ckw}. 

LHCb reported the first measurement of {\it CP} violation in $B\rightarrow D^{*\pm}D^{\mp}$ at LHCb using the full Run~1 and Run~2 data set, corresponding to an integrated luminosity of $9$ $\mathrm{fb^{-1}}$~\cite{betaDstarD}. These decays are mediated by a $b\to c\overline{c}d$ transition sensitive to $\beta$. In addition, a penguin contribution at the level of a few percent is expected. Therefore, a comparison of time-dependent asymmetries in $B\to D^{*\pm}D^{\mp}$ with those measured in $b\to c\overline{c}s$ transitions is a probe for physics beyond the Standard Model. LHCb measures $S(D^{*}D) = -0.861 \pm 0.077\text{(stat)} \pm 0.019 \text{(syst)}$, which is compatible with the LHCb combination in $b\rightarrow [c\overline{c}]s$ decays, and the results of all {\it CP}-violating parameters are in a good agreement with previous measurements of the B-factories~\cite{betaDstarDBaBar,betaDstarDBelle}. The precision of $\Delta C(D^*D)$  and $C(D^*D)$ is comparable with that of previous measurements, while for $S(D^*D)$, $\Delta S(D^*D)$ and $A(D^*D)$, this measurement improve significantly the
precision of the current world average~\cite{Amhis:2019ckw}. The measurement excludes the hypothesis of {\it CP} conservation at more than 10$\sigma$.

\subsection{Measurement of the {\it CP}-violating phase $\phi_{s}$ at the LHC}

\begin{figure}[h!]
     \centering
     \includegraphics[width=0.6\linewidth]{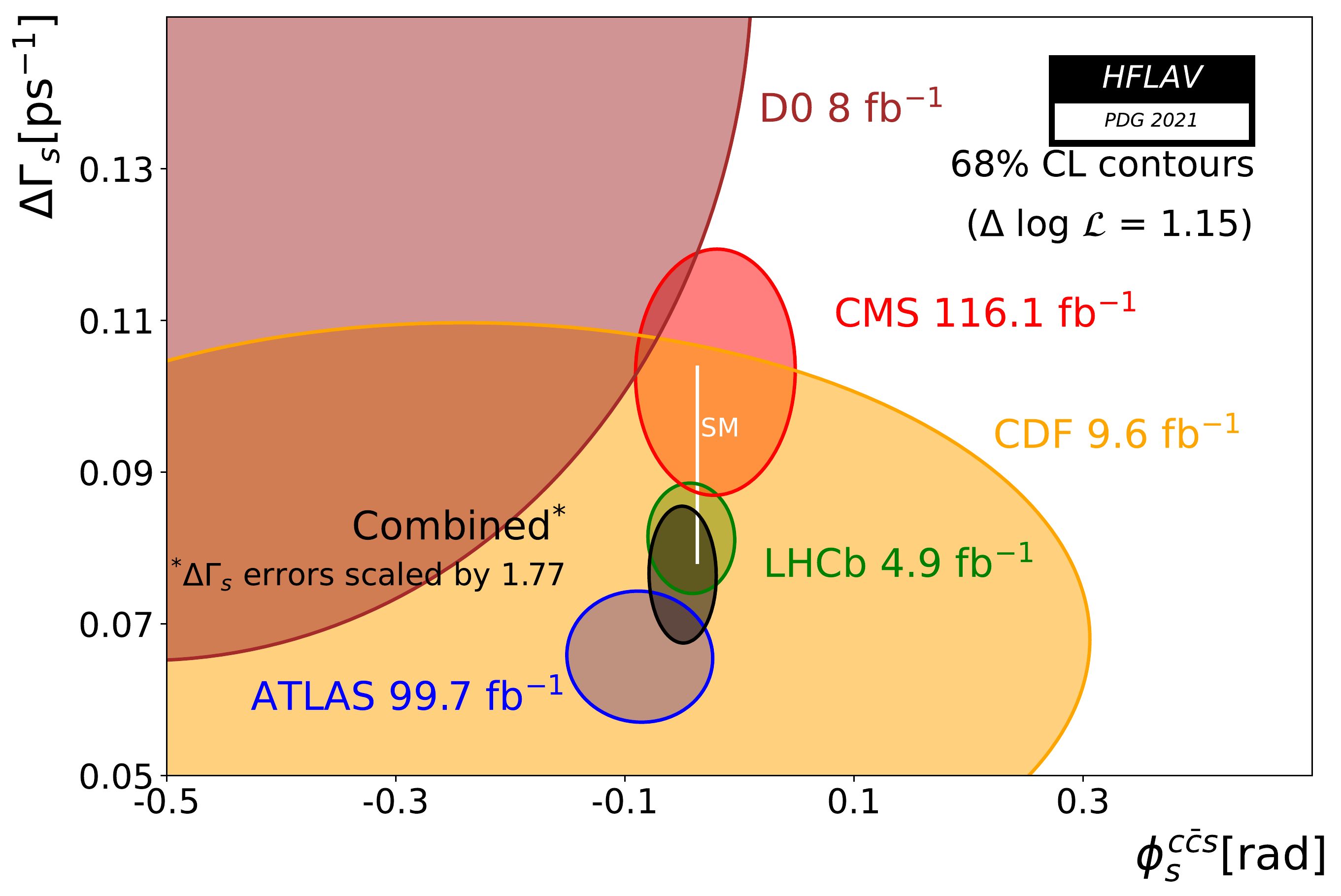}
     \caption{Measurements of $\phi_s$ and $\Delta\Gamma_s$ with individual 68\% confidence-level contours from measurements of ATLAS~\cite{phis-atlas-2014,phis-atlas-2016,phis-atlas}, CDF~\cite{phis-cdf},  CMS~\cite{phis-cms-2015,phis-cms}, D0~\cite{phis-d0} and LHCb~\cite{phis-2015-LHCb,phis1,phis2,phis3,phis4,phis5} and the combined contour (black solid line and shaded area), as well as the Standard Model predictions (white rectangle)~\cite{CKMfitter,UTfit:2007eik,LenzDGs}.}
     \label{wa}
\end{figure}

The ATLAS, CMS and LHCb experiments reported measurements of the {\it CP}-violating phase $\phi_{s}$ and lifetime and mixing  observables in the decay mode $B_{s}^{0}\to J/\psi(\mu\mu) \phi$~\cite{phis-atlas,phis-cms,phis1} using Run~2 LHC data. The results in $\phi_{s}$ are in a good agreement while the time and angular observables exhibit tensions at the level of several $\sigma$.
 The world averages of these and previous $\phi_{s}$ measurements~\cite{phis1,phis2,phis3,phis4,phis5,phis-2015-LHCb,phis-atlas-2014,phis-atlas-2016,phis-atlas,phis-cms-2015,phis-cms,phis-d0,phis-cdf} are $\phi_{s}=-50\pm19$~rad and $\Delta\Gamma_{s}=0.077\pm0.006$~ps$^{-1}$~\cite{Amhis:2019ckw} and are illustrated in Fig.~\ref{wa}.

A first measurement of $B_{s}^{0}\to J/\psi(e^{+}e^{-}) \phi$ based on Run~1 data was reported by LHCb~\cite{electron}. The yield of the $B_s^0 \rightarrow J/\psi(e^+e^-)\phi$ sample corresponds to about $10\%$ of the  $B_s^0 \rightarrow J/\psi(\mu^+\mu^-)\phi$ mode~\cite{phis-2015-LHCb}. The results, $\phi_s= 0.00 \pm 0.28 \pm 0.07$ rad, $\Delta\Gamma_s = 0.115 \pm 0.045 \pm 0.011$ $\mathrm{ps^{-1}}$ and $\Gamma_s = 0.608 \pm 0.018 \pm 0.012$ $\mathrm{ps^{-1}}$ are consistent with previous measurements~\cite{phis1,phis5}, SM predictions from global fits to experimental data~\cite{CKMfitter,UTfit:2007eik} and show no evidence of {\it CP} violation in the interference between $B_s^0$ meson mixing and decay.

\subsection[Penguin effects in BdJpsiKs and BsJpsiPhi]{\boldmath Penguin effects in $B_d^0\to J/\psi K^0_{\text{S}}$ and $B_s^0\to J/\psi\phi$}
The discovery of New Physics contributions to the {\it CP}-violating phases $\phi_d$ and $\phi_s$, associated with mixing between neutral $B_q^0$ and $\overline B_q^0$ mesons $(q=d,s)$, relies both on improved experimental measurements and on equally small theoretical uncertainties associated with the interpretation of these results.
To achieve the latter, it is necessary to control contributions from higher-order decay topologies, which are often still neglected today, in all the decay channels that are used to measure $\phi_d$ and $\phi_s$.
In particular, this applies to the doubly Cabibbo-suppressed penguin topologies affecting the decay channels $B_d^0\to J/\psi K^0_{\text{S}}$ and $B_s^0\to J/\psi\phi$, which are considered the golden modes for the determination of $\phi_d$ and $\phi_s$, respectively.
Due to the presence of these penguin topologies, the {\it CP} asymmetries in $B_d^0\to J/\psi K^0_{\text{S}}$ and $B_s^0\to J/\psi\phi$ only allow us to measure effective mixing phases $\phi_q^{\text{eff}}$, which are related to $\phi_q$ via hadronic shifts $\Delta\phi_q$.
The $\Delta\phi_q$ are of the same order as the current experimental uncertainties to $\phi_q^{\text{eff}}$ and thus will become the dominant sources of systematic uncertainty in the determination of $\phi_d$ and $\phi_s$ if penguin effects remain unaccounted for.

The penguin shifts $\Delta\phi_q$ can be determined with a strategy employing the $SU(3)$ flavour symmetry of QCD, as discussed in Ref.\ \cite{Barel:2020jvf}.
At the CKM 2021 conference, updated results from this analysis were presented.
From a simultaneous analysis of the decays $B_d^0\to J/\psi K^0_{\text{S}}$ and $B_s^0\to J/\psi\phi$ and their penguin control modes $B_s^0\to J/\psi K^0_{\text{S}}$, $B_d^0\to J/\psi\pi^0$ and $B_d^0\to J/\psi\rho^0$, we find
\begin{equation}
    \phi_d = \left(44.4_{-1.5}^{+1.6}\right)^{\circ}\:,\qquad
    \phi_s = -0.074_{-0.024}^{+0.025} = \left(-4.2 \pm 1.4\right)^{\circ}\:.
\end{equation}
Comparing these fit values with the experimental inputs
\begin{equation}
    \phi_{d,J/\psi K^0}^{\text{eff}} = \left(43.6 \pm 1.4\right)^{\circ}\:,\qquad
    \phi_{s,J/\psi\phi}^{\text{eff}} = -0.071 \pm 0.022 = (-4.1 \pm 1.3)^{\circ}
\end{equation}
clearly shows the small, but non-negligible impact of the penguin topologies in these decays.

\section{Time-dependent measurements at Belle and Belle II}

Measurements of decay-time dependent {\it CP} violation with $B^0$ mesons are central in the physics program of Belle and of its
upgrade Belle II.
In particular, mixing-induced {\it CP} violation in tree-level $B^0\rightarrow J/\psi K^0_S$ decays gives access
to the CKM angle $\beta$. Belle II aims at reducing the uncertainty on $\beta$ by a factor $\sim 5$ with respect to the
current world average to reach a precision of $\sim0.2\deg$~\cite{Belle-II:2018jsg}. In addition to this precision measurement,
time-dependent {\it CP} violation is used as a probe for New Physics in rare, penguin mediated, $B^0$ decay.

One example of a penguin mediated transition is the $B^0\rightarrow K_S^0 K_S^0 K_S^0$ decay. The SM predicts no direct {\it CP} asymmetry
in this decay, $\mathcal{A}=0$, and the time-dependent {\it CP} violation parameter is expected to be $\mathcal{S}=-\sin2\beta$.
The time-dependent analysis of this decay is performed using the full Belle dataset, corresponding to $711\;\text{fb}^{-1}$ of
$e^+e^-$ collision data at the $\Upsilon(4S)$ resonance.
%One challenge of this analysis is the control of the background stemming from pairs of light quarks. This background 
%is reduced using a neural network trained on event shape variables.
The clean environment of Belle is especially suited for this measurement, as all tracks coming from the $K^0_S$ mesons are detached from the
$B^0$ decay vertex.
Therefore, to measure the position of the $B^0$ decay vertex, needed to compute the decay time, the intersection between the lines of flight
of the $K^0_S$ mesons and a constraint constructed from the known position of the $e^+e^-$ collision point is used.
The analysis finds $\mathcal{S}=-0.71\pm0.23\,(\text{stat.})\pm0.05\,(\text{syst.})$ and
$\mathcal{A}=0.12\pm0.16\,(\text{stat.})\pm0.05\,(\text{syst.})$~\cite{Belle:2020cio}, compatible with the SM expectation.

In preparation for precision analyses, the Belle II collaboration has performed several measurements
illustrating the nominal performance of the detector for time-dependent studies. Using $34\;\text{fb}^{-1}$ of data collected until summer
2020, the time-dependent {\it CP}-violation parameter in $B^0\rightarrow J/\psi K^0_S$ decays is measured to be
$\mathcal{S}=0.55\pm0.21\,(\text{stat.})\pm0.04\,(\text{syst.})$.
Using the same data and $B^0\rightarrow D^-\pi^+$ decays, the
$B^0-\overline{B}^0$ oscillation frequency is measured to be $\Delta m_d = 0.531\pm0.046\,(\text{stat.})\pm0.013\,(\text{syst.})\;\text{ps}^{-1}$,
both compatible with the world average. Using $63\;\text{fb}^{-1}$ and hadronic $B^0$ channel, a time integrated analysis is performed to
extract the effective tagging efficiency $\varepsilon_{\text{tag}}=(30.0\pm1.3)\%$~\cite{Belle-II:2021zvj}.
The performance of the Belle II flavour tagger is hence already comparable to the Belle one ($\varepsilon_{\text{tag}}=(30.1\pm0.4)\%$~\cite{BaBar:2014omp}) at
this early stage of data taking. % and is expected to improve significantly in the future.  
Finally, using $72\;\text{fb}^{-1}$ of data, 
the Belle II collaboration provides the most precise measurement of the $D^0$ and $D^+$ lifetimes to date:  
$\tau(D^0)=410.5\pm1.1\,(\text{stat.})\pm0.8\,(\text{syst.})\,\text{fs}$ and $\tau(D^+)=1030.4\pm4.7\,(\text{stat.})\pm3.1\,(\text{syst.})\,\text{fs}$.
This high level of precision illustrates the good performance of the upgraded vertex detector, including the new pixel detector
situated $1.4\;\text{cm}$ from the $e^+ e^-$ interaction region, and of the accuracy of the alignment of the tracking system. 

\section{Puzzles in the $B^0_s\to D_s^\mp K^\pm$ system}
The $B^0_s\to D_s^\mp K^\pm$ system, consisting of pure tree decays, is particularly interesting for testing the SM description of {\it CP} violation \cite{ADK,RF-BsDsK,DeBFKMST}. An intriguing value of the angle $\gamma$ of the Unitarity Triangle (UT) was reported by LHCb \cite{LHCb-BsDsK}. In order to gain a better understanding of this result, Malami and collaborator Fleischer have performed a transparent analysis of the corresponding {\it CP} asymmetries, obtaining the value of $\gamma=\left(131^{+17}_{-22}\right)^\circ$  \cite{Fleischer:2021cct,Fleischer:2021cwb}, which is in excellent agreement with the LHCb picture. Here, they have paid special attention to discrete ambiguities, resolving a remaining final one and using a value of the $B_{s}^0-\overline{B}_{s}^0$ mixing phase $\phi_s$, which includes penguin corrections. This surprisingly large result is in tension with global analyses of the UT, which give values around $70^\circ$ \cite{Amhis:2019ckw,PDG}. 

Complementing the {\it CP}-violating observables with information from branching ratios, one arrives at another puzzling situation. The individual branching ratios of the two decay channels are first determined from the data. For the theoretical SM interpretation, these are converted into effective colour factors $|a_1|$, characterising colour-allowed tree decays. To this end, information from $B_{(s)}$ semileptonic decays is utilised, allowing the extraction of these parameters in the cleanest possible way with respect to uncertainties from CKM parameters and hadronic form factors. A prime example, where QCD factorisation \cite{Beneke:2000ry} is expected to work excellently, is the $\overline{B}^0_s \rightarrow D_s^+ K^-$ channel. One finds that additional contributions from exchange topologies, which are non-factorisable, play a minor role and do not indicate any anomalous behaviour. A surprisingly small value of $|a_1|$ is obtained, which is in tension with the theoretical prediction \cite{Huber:2016xod}. A similar pattern arises in the $\overline{B}^0_s \rightarrow K^+D_s^-$ channel. Applying this method also to other $B_{(s)}$ decays with similar dynamics to complement the analysis, one arrives at consistent results, thereby making the intriguing situation even more exciting.

In view of these puzzles, Fleischer and Malami have developed a model-independent formalism, generalising the analysis of the $B^0_s\to D_s^\mp K^\pm$ system to include NP effects. They go beyond assumptions made in the LHCb analysis~\cite{LHCb-BsDsK} and apply their strategy to the current data. This allows the calculation of correlations between the NP parameters and their {\it CP}-violating phases. They find that they can describe both the {\it CP} violation and the branching ratio measurements with NP contributions at the level of $30\%$ of the SM amplitudes. This strategy can be fully exploited in the future high precision era of $B$ physics. It is exciting to see whether the tantalising question can be answered: Could new sources of {\it CP} violation be established?

\section{New ideas for $\phi_2$ ($\alpha$) measurements}

Although $\phi_2$ has become the least known experimental input to fits of the Unitarity Triangle, a number of technical challenges may limit potential improvements of its precision if not addressed. For the record, measurements are strictly sensitive to the weak phase $-2\phi_1 - 2 \phi_3$ ($-2\beta - 2\gamma$). While there is no practical difference when constraining the Standard Model, the distinction does become important for example, when parameterising possible New Physics contributions in neutral $B - \overline B$ mixing. Four analysis strategies are proposed to improve the interpretation of $\phi_2$. 

Firstly, the analysis space of $B^0 \to \rho^0 \rho^0$ can be increased to include interfering $B^0 \to a_1^\pm \pi^\mp$ decays in a time-dependent flavour-tagged amplitude analysis. The known penguin contamination in $B^0 \to a_1^\pm \pi^\mp$, will prevent the $C\!P$-violating parameter of each amplitude contribution from factorising in the isobar sum, thereby allowing an effective $\phi_2$ to be determined unambiguously within the range $[0,180]^\circ$~\cite{Dalseno:2018hvf}. Consequently, a single $\phi_2$ solution as constrained from the SU(2) isospin analysis~\cite{Gronau:1990ka} is also resolved.

With the approach described above, the effective weak phases of $B^0 \to a_1^\pm \pi^\mp$ will also be resolved without ambiguity in the range $[0,180]^\circ$. An amplitude analysis of the $B^+ \to K_S^0 \pi^+ \pi^- \pi^+$ is proposed to determine the complex couplings of the $B^+ \to K_{1A}^0 \pi^+$ and $K^0 a_1^+$ contributions, the $K_{1A}$ being the ${}^3P_1$ partner of the $a_1$. In a subsequent SU(3) analysis, the number of $\phi_2$ solutions is reduced from 8~\cite{Gronau:2005kw} down to 1~\cite{Dalseno:2019kps}. More importantly, these amplitude analyses also offer sufficient degrees of freedom to constrain non-factorisable SU(3)-breaking effects for a precision measurement of $\phi_2$, providing a consensus is eventually achieved on the $K_1$ mixing angle.

Given that the dipion masses are modelled in any analysis of the $B \to \rho \rho$ system, a new source of $\phi_2$ bias is identified in addition to those already established in $B \to \rho \rho$, including $\rho^0-\omega$ mixing~\cite{Gronau:2005pq} and the finite $\rho$ width~\cite{Falk:2003uq}. A lack of coordination in how systematic uncertainties arising from the $\rho$ pole properties are propagated can also lead to a non-negligible bias~\cite{Dalseno:2021bin}. This effect can even be exacerbated in combination with the $\phi_2$ measurement coming from $B^0 \to (\rho \pi)^0$. By applying the same systematic variations to all measurements containing a $\rho$ mesons, the systematic covariance matrix that ensues will eliminate such bias in the $\phi_2$ combination.

In the final improvement, a rescaling of the SU(2) isospin triangles is suggested by dividing through by the base length~\cite{Dalseno:2021vgq}, which can be considered a nuisance parameter in the pursuit of $\phi_2$. The isospin triangles would then be constrained by ratios of branching fractions instead of their absolute measurements. As ratios are much cleaner experimentally through the cancellation of several systematic uncertainties, this approach paves the way to a more systematically sustainable $\phi_2$ analysis. Interestingly, this approach also opens the possibility for LHCb to make an independent measurement of $\phi_2$, exploiting a peculiarity in the $B \to \rho \rho$ triangle geometry. As these triangles are known to be essentially flat, a meaningful constraint of $\phi_2$ can nevertheless be achieved without the need for a time-dependent flavour-tagged analysis of $B^0 \to \rho^+ \rho^-$, with its measured yield being as small as a few hundred events.

\section{Global fits to the Unitarity Triangle}
Here, we summarise the contributions on the latest status of UTFit and CKMFitter.

\subsection{Updates in the Unitarity Triangle fits with UTfit}
Flavour physics provides some of the most stringent tests of the SM. In the last two decades, the \textbf{UT}\emph{fit} collaboration has regularly provided updates of the Unitarity Triangle (UT) analysis, constantly improving the knowledge of the CKM matrix parameters~\cite{Cabibbo:1963yz,Kobayashi:1973fv}. The UT triangle fit~\cite{UTfit:2006vpt} has been performed in two different scenarios. The first is a SM analysis aiming to make comparisons with SM predictions and assess their compatibility. The latter is a NP analysis in which the most generic NP loops are added to the SM structure to probe their contribution to $\Delta F = 2$ tranistions.

\begin{figure}[!htb]                                                    \centering 
    \includegraphics[width=0.4\textwidth]{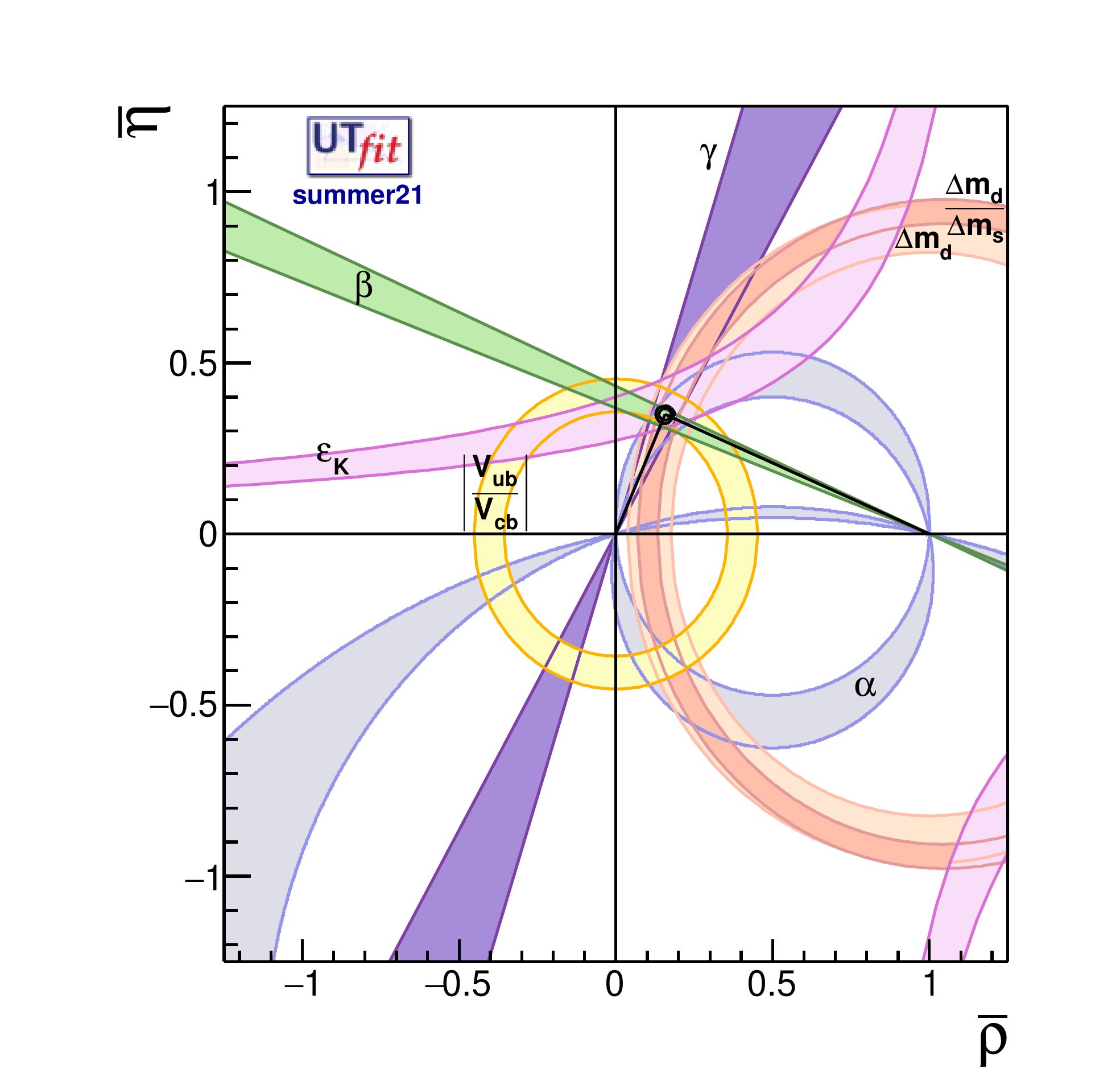}
    \includegraphics[width=0.4\textwidth]{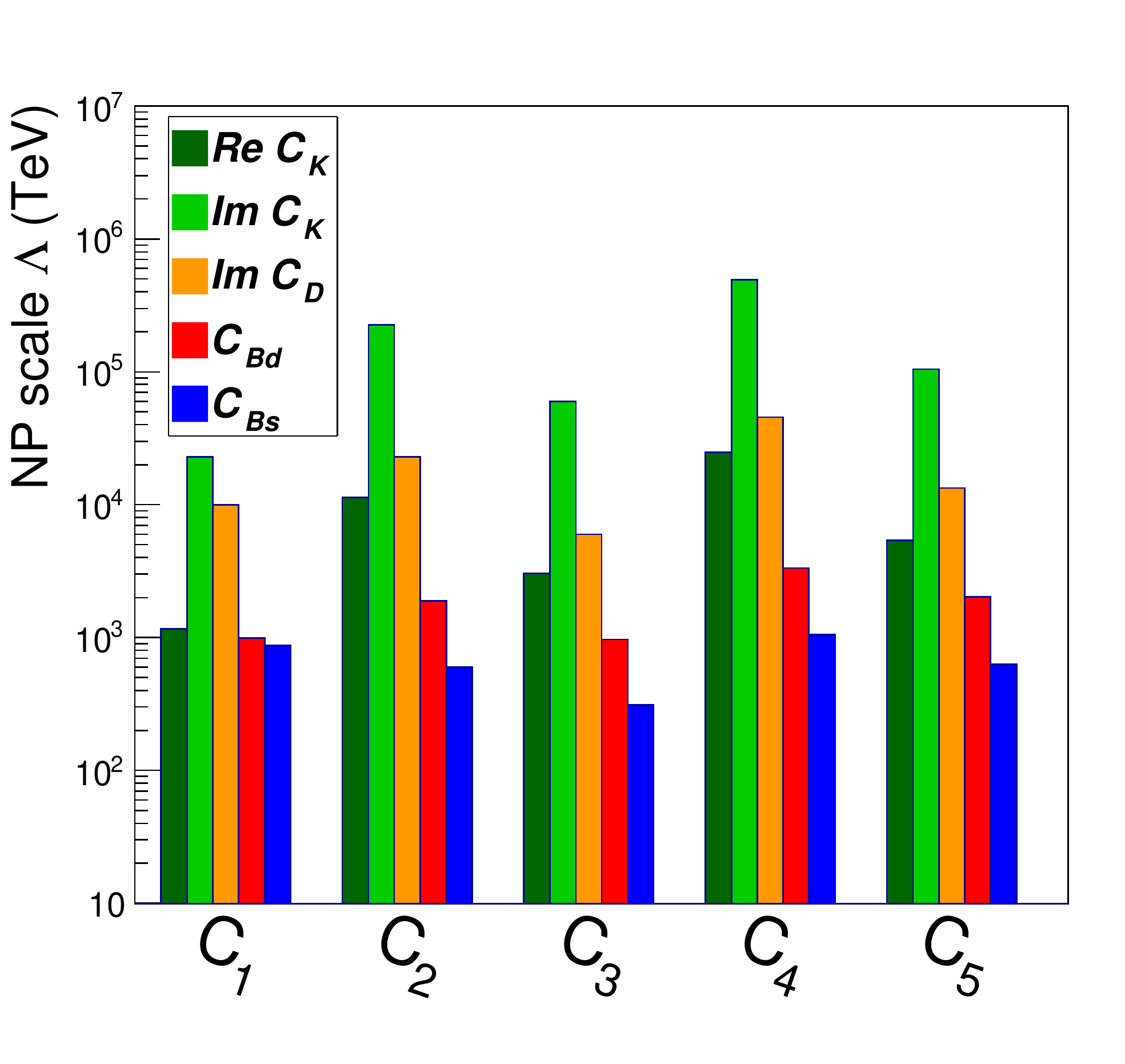}            \caption{(Left) $\overline{\eta}-\overline{\rho}$ plane with the results of the SM fit for the UT apex. The various constraints and allowed regions $(95\%\ \rm{probability})$ are also shown. (Right) Summary of the $95\%$ probability lower bound on the NP scale $\Lambda$ for a generic NP scenario.}
   \label{fig:UTfit}                
\end{figure}

Using the most up-to-date experimental, phenomenological and LQCD inputs within the \textbf{UT}\emph{fit} Bayesian framework, a global fit is performed to determine the CKM matrix parameters $\overline{\rho}$ and $\overline{\eta}$. Their values are found to be $0.155 \pm 0.011$ and $0.350 \pm 0.010$, respectively. The SM fit results in the $\overline{\rho} - \overline{\eta}$ plane are shown in the left part of Fig.~\ref{fig:UTfit}. This analysis shows a very good compatibility with the SM, but the historical tension between the inclusive and exclusive determinations of $|V_{\rm ub}|$ and $|V_{\rm cb}|$ is still present and more data will be needed to clarify the picture.

The NP analysis exploits additional inputs to search for contributions beyond the SM. The results show a good compatibility with the SM expectations, but NP contributions at the $10-20\%$ level are still allowed. The results can also be translated into allowed ranges for the Wilson coefficients of the effective Hamiltonian~\cite{UTfit:2007eik}. For example, by considering a generic strongly interacting theory with arbitrary flavour structure one can obtain lower bounds for the NP scale $(\Lambda)$. They are shown for different Wilson coefficients in the right part of Fig.~\ref{fig:UTfit}. The strongest bound on the NP scale comes from $\rm{Im}\ C_K$ of the fourth coefficient and corresponds to $\Lambda > 4.3 \cdot 10^5$~TeV. This confirms the great power of flavour physics in imposing constraints on quantities not currently reachable with direct searches.

\subsection{Updates on global fits from the CKMfitter group}
The CKMfitter group aims at performing global fits of the CKM parameters by combining efforts from both experimental and theoretical sides. In the CKMfitter, a frequentist approach based on a $\chi^2$ analysis is used to combine different measurements and lattice inputs. The Range fit (Rfit) scheme is used to treat statistical and theoretical uncertainties, where different sources of uncertainties from theoretical inputs are summed linearly and the final uncertainty due to theoretical assumptions is treated as a range, instead of a Gaussian distribution~\cite{Rfit}. The statistical uncertainty is still considered as Gaussian like, where the central values are those determined by the range.

\begin{figure}
	\centering
	\caption{Global fit results of the CKM parameters shown on $\bar{\rho}$-$\bar{\eta}$ plane. Filled areas correspond to 95\% Confidence Level.}\label{fg:gfit}
		\includegraphics[width=7.5cm]{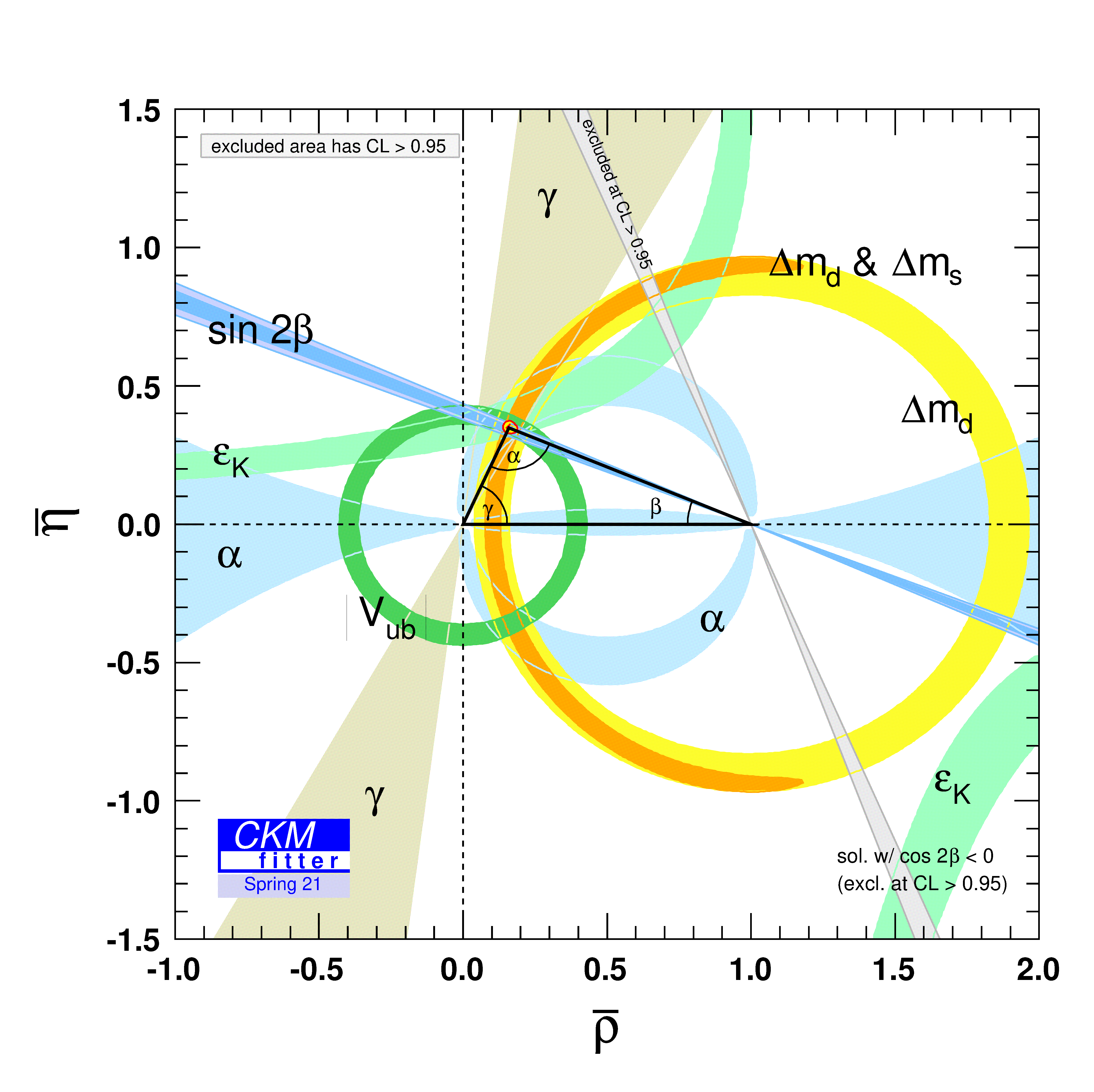}
\end{figure}

The latest CKM fit results with inputs till early 2021 (Moriond 2021) are shown in Fig.~\ref{fg:gfit}. The $\chi^2_{\textrm{min}}$, corresponding to a $p$-value of 29\%, is increased slightly compared to the 2019 results~\cite{CKMfitter}. The Wolfenstein parameters are determined to be 
\begin{eqnarray}
A &=& 0.8132^{+0.0119}_{-0.0060}, \qquad  \lambda = 0.22500^{+0.00024}_{-0.00022}, \\\nonumber
\bar{\rho} &=& 0.1566^{+0.0085}_{-0.0048}, \qquad \bar{\eta} = 0.3475^{+0.0118}_{-0.0054},
\end{eqnarray}
and the Jarlskog invariant to be $J = (3.044^{+0.068}_{-0.084})\times 10^{-5}$. 
The global fits are also performed using different sets of selected observables, such as {\it CP} violation only or {\it CP} conserving only observables,  observables determined only from tree-level processes or with loop-level processes involved etc., all show consistent pictures. More results can be obtained from the \href{http://ckmfitter.in2p3.fr}{CKMfitter webpage}.

\section{New physics in $ B $ meson mixing: future sensitivity and limitations}

\begin{figure*}[t]
\includegraphics[scale=0.38,clip,bb=15 15 550 470]{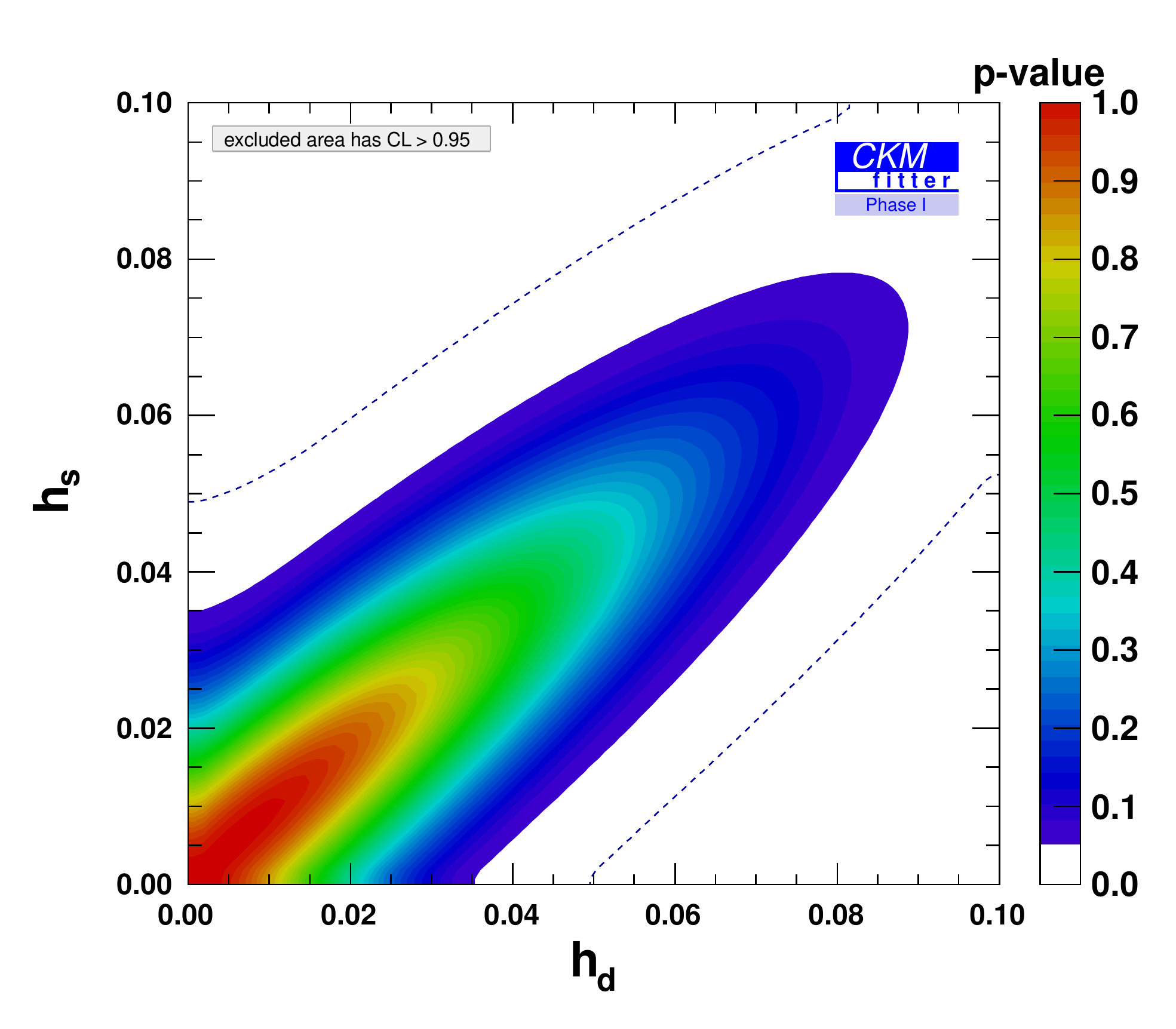}
\hfil
\includegraphics[scale=0.38,clip,bb=15 15 550 470]{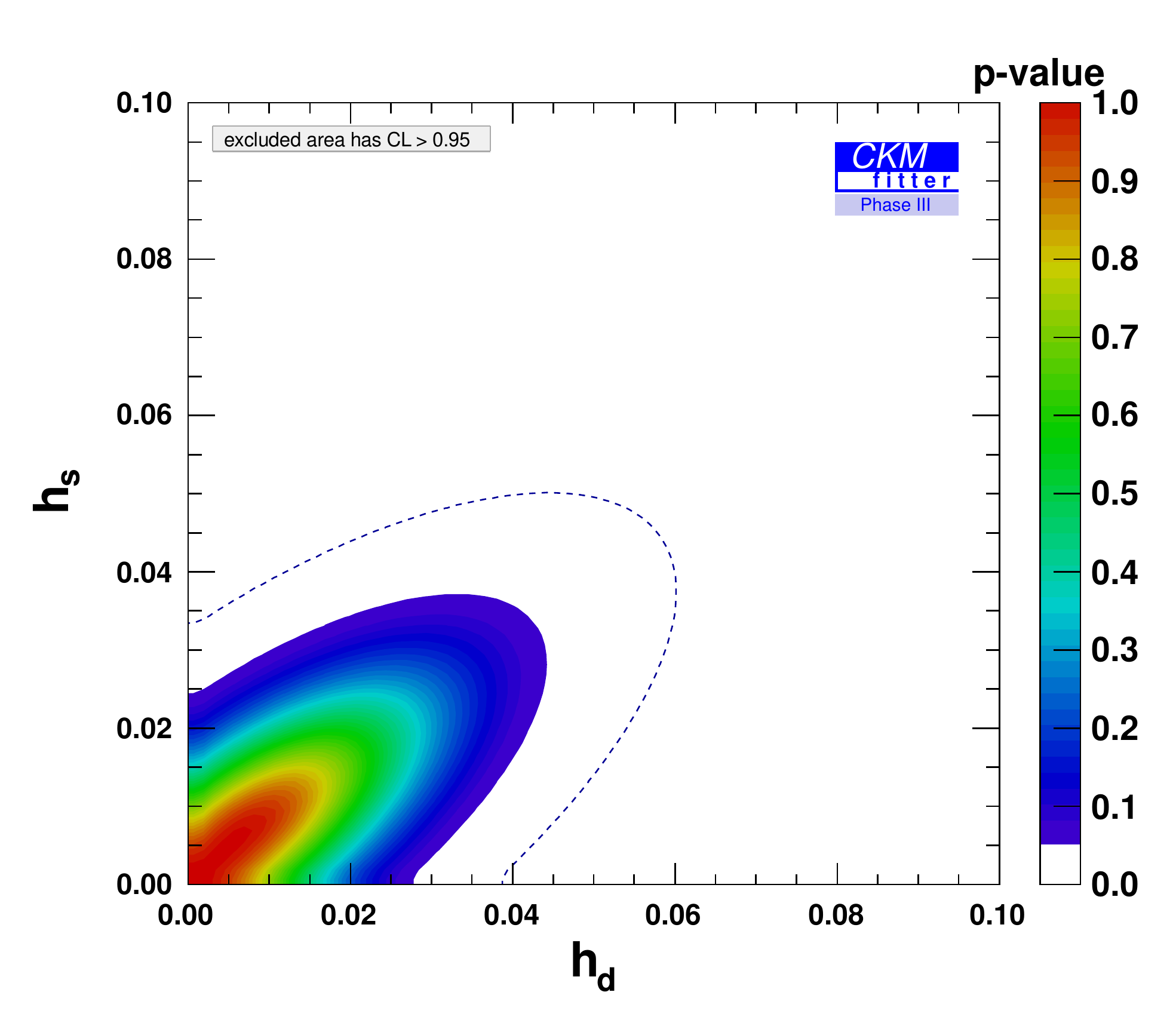}
\caption{Phase~I (left) and Phase~III (right) sensitivities to $h_d-h_s$ in $B_{d}$ and $B_{s}$ mixings.
The SM point corresponds to the origin.
The dotted curves show the 99.7\%\,CL ($3\sigma$) contours.
}
\label{fig:hdhs_plots}
\end{figure*}

The planned LHCb Upgrades, Belle~II and its possible upgrade, and the tera-$Z$ phase of the proposed FCC-$e e$ program have a huge potential of unveiling New Physics (NP) contributions affecting flavour observables \cite{Cerri:2018ypt,Belle-II:2018jsg,FCC:2018byv}.
This is illustrated for neutral meson mixing in Fig.~\ref{fig:hdhs_plots}, where $ h_d $ and $ h_s $ parametrize the sizes of NP contributions relative to the SM. While presently the bounds on NP are worse than the scale of the plots, near future bounds achievable by the end of this decade will push these contributions below the 10\% level (@ 95\% CL), see the left panel of Fig.~\ref{fig:hdhs_plots} (Phase~I: LHCb 50/fb $+$ Belle~II 50/ab).
The combination LHCb 300/fb $+$ Belle~II 250/ab (named Phase~II) is expected to lead to a less impressive improvement of these constraints, by a factor 1.5 compared to Phase~I.
To increase the latter factor, progress is needed in key quantities beyond current expectations, namely, the determinations of hadronic inputs (decay constants and bag parameters) and perturbative QCD corrections, and the extraction of the CKM matrix element $ |V_{cb}| $. Improving the latter by a sizeable factor (namely, 20) has an effect similar to the one achievable by adding FCC-$e e$ to Phase~II, seen in the right panel of Fig.~\ref{fig:hdhs_plots} (Phase~III: Phase~II $+$ FCC-$e e$ tera-$Z$), which however results from a first look into FCC-$e e$ flavour physics capabilities.
These constraints on $h_d-h_s$ translate into sensitivities to tree level NP contributions to meson mixing at the scale of hundreds ($B_s$ mixing) to thousands ($B_d$ mixing) of TeV.
More details are found in \cite{Charles:2020dfl} and in a dedicated article in these conference proceedings.

\section{Summary}
Significant advances in the precision of mixing and mixing-related {\it CP} observables, both in experiment and in theory, were presented at CKM 2021. These provide important insights into the deeper understanding of the fundamental principles of nature.

The experimental precision in $\Delta m_{s}$ was improved by LHCb by more than a factor of three compared to CKM 2018. The global precision on $\phi_{s}$ was improved by over 30\% with the latest ATLAS, CMS and LHCb $B^{0}_{s}\to J/\psi KK$ measurements, requiring a deeper understanding of the underlying penguin contributions. 
At the same time, lattice QCD continues to reduce uncertainties in hadronic matrix elements, and higher-order contributions in the heavy quark expansion are being calculated.
In a global fit to $B^{0}_{s}\to J/\psi KK$, $B^{0}_{s}\to J/\psi K_{\rm S}^{0}$ and control modes, the corresponding penguin contributions to $\phi_{s}$ and $\phi_{d}$ were shown to be small yet non-negligible. LHCb measured the time-dependent {\it CP} asymmetry in the $B_{s}^{0}$ system for the first time, in the $B_{s}^{0}\to KK$ decay. Results from Belle~II early data, such as the world's most precise measurement of $\tau(D^{0})$ and $\tau(D^{+})$, illustrate the good performance of the vertex detector, including the new pixel detector.
The latest measurements as well as the global fits to experimental data agree with the Standard Model predictions. Nevertheless, the increase in precision is an important step towards constraining contributions beyond the Standard Model.
Finally, to optimally exploit current and forthcoming data, new methods are proposed that pave the way for more systematically sustainable $\alpha(\phi_{2})$ analyses and further constrain NP contributions in $B_{s}^{0}\to D_{s}^{\mp}K^{\pm}$.
\section*{Acknowledgements}

V.\,S. is very grateful to Marvin Gerlach, Matthias Steinhauser and Ulrich Nierste for the collaboration on the project reported in Sec.\,\ref{sec:Bmixing_theory}. His research was supported by the Deutsche Forschungsgemeinschaft (DFG, German Research Foundation) under grant 396021762 --- TRR 257 ``Particle Physics Phenomenology after the Higgs Discovery''. This contribution is registered under the preprint numbers
P3H-22-020 and TTP22-013.
The project of L.\,V.\,S. has received funding from the European Union’s Horizon 2020 research and innovation programme under the Marie Sklodowska-Curie grant agreement No 101031558; his work has also been supported in part by MCIN/AEI/10.13039/501100011033 Grant No. PID2020-114473GB-I00, by PROMETEO/2021/071 (GV).
J.T.T. acknowledges that the project leading to this application has received funding from the European Union's Horizon 2020 research and innovation programme under the Marie Sk{\l}odowska-Curie grant agreement No 894103.

\end{document}